\documentstyle[aps,prl,preprint,epsf,floats]{revtex}

\tightenlines

\newcommand{\beq}{\begin{equation}}
\newcommand{\eeq}{\end{equation}}
\newcommand{\bqa}{\begin{eqnarray}}
\newcommand{\eqa}{\end{eqnarray}}
\def\sumint{\hbox{$\sum$}\!\!\!\!\!\!\int}
\def\sumintsmall{\hbox{$\sum$}\!\!\!\!\!{\displaystyle \int}}

\begin{document}

\title{Mass Expansions of Screened Perturbation Theory}
\preprint{
\vbox{\halign{&##\hfil\cr
&ITF-UU-01/19 \cr
&NT@UW-01-11 \cr
        & hep-ph/yymmnn \cr
&\today\cr }}}

\author{Jens O. Andersen
}
\address{ITP, University of Utrecht,
       Leuvenlaan 4, 3584 CE Utrecht, The Netherlands}

\author{Michael Strickland}
\address{Physics Department, University of Washington, Seattle  WA 98195-1560}

\maketitle
\begin{abstract}
The thermodynamics of massless $\phi^4$-theory is studied within screened
perturbation theory (SPT). In this method the perturbative expansion
is reorganized by adding and subtracting a mass term in the Lagrangian.
We analytically calculate the pressure and entropy to
three-loop order and the screening mass to two-loop order, 
expanding in powers of $m/T$. 
The truncated $m/T$-expansion results are compared with numerical SPT results for 
the pressure, entropy and screening mass which are accurate to all orders in $m/T$.
It is shown that the $m/T$-expansion converges quickly and provides an 
accurate description of the thermodynamic functions for large values
of the coupling constant.
\end{abstract}

\newpage
\section{Introduction}

The behavior of finite temperature field theory at intermediate 
to large coupling is of particular interest due to the upcoming
heavy-ion collision experiments at RHIC and LHC.  For
years the hope has been that due to the asymptotic freedom of
QCD that weak-coupling expansion calculations within finite
temperature field theory would suffice to describe the experimental
data.  Along these lines, there has been significant 
progress in recent years in perturbative calculations within thermal 
field theory.  The pressure in QCD, for example, is now known to
order $g^5$ \cite{arnold-zhai,Kastening-Zhai,Braaten-Nieto:QCD}.
Unfortunately, an analysis of the convergence of this expansion
shows that the successive perturbative approximations do not 
converge for experimentally accessible temperatures.  
This lack of convergence, while not surprising, needs to be addressed
in order to provide systematic methods for calculating quark-gluon
plasma observables.

\begin{figure}[t]
\epsfysize=8cm
\centerline{\epsffile{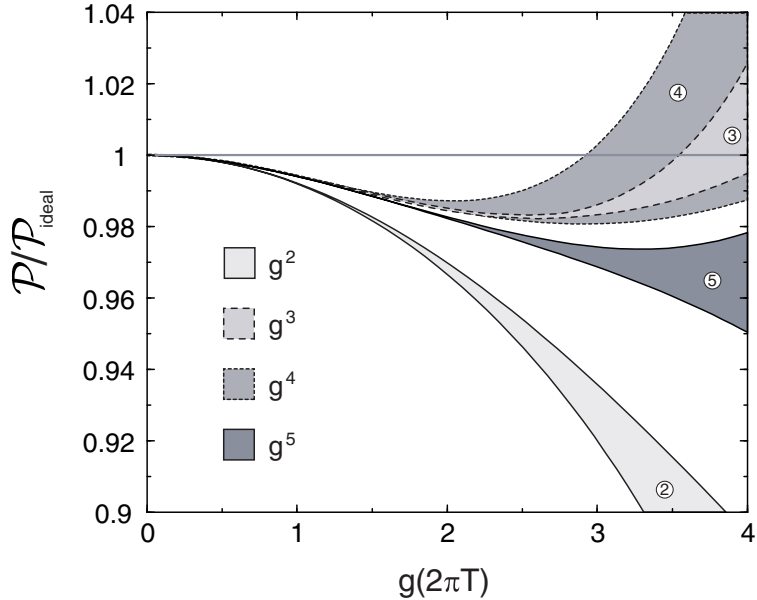}}
\vspace{3mm}
\caption[a]{Weak-coupling expansion to orders $g^2$, $g^3$, $g^4$, and $g^5$
for the pressure normalized to that of an ideal gas as a function
of $g(2\pi T)$.}
\label{fpert}
\end{figure}

The lack of convergence of the weak-coupling expansion is not restricted
to QCD.  In fact, even in simple massless scalar field theories similar
convergence problems are encountered.  This indicates that the problem
might be universal.  The universality of the problem means that the technique
needed might be quite general and since calculations within scalar theories are
technically simpler than in full QCD these theories can provide an important 
testing ground for methods to deal with this problem.  
Like QCD, the weak-coupling expansion for the pressure of 
a massless scalar field theory with a $g^2\phi^4/4!$ interaction, 
is known to order $g^5$~\cite{arnold-zhai,Parwani-Singh,Braaten-Nieto:scalar}
\bqa\nonumber
{\cal P} &=& {\cal P}_{\rm ideal} \left[
1-{5\over4}\alpha+{5\sqrt{6}\over3}\alpha^{3/2}+{15\over4}
\left(\log{\mu\over2\pi T}+0.40\right)\alpha^2\right.\\ 
&&\left.-{15\sqrt{6}\over2}\left(\log{\mu\over2\pi T}-{2\over3}\log\alpha
-0.72\right)\alpha^{5/2}+{\cal O}(\alpha^3\log\alpha)\right]\;,
\eqa
where ${\cal P}_{\rm ideal} = (\pi^2/90)T^4$
is the pressure of an ideal gas of free massless bosons,
$\alpha=g^2(\mu)/16\pi^2$, and $g(\mu)$ is the 
$\overline{\rm MS}$ coupling constant at the renormalization scale $\mu$.
In Fig.~\ref{fpert}, we show the successive perturbative approximations to
${\cal P}/{\cal P}_{\rm ideal}$ as a function of $g(2\pi T)$. Each partial
sum is shown as a band obtained by varying $\mu$ 
from $\pi T$ to $4\pi T$.
To express $g(\mu)$ in terms of
$g(2\pi T)$, we use the numerical 
solution to the renormalization group equation
$\mu{\partial\over \partial\mu}\alpha=\beta(\alpha)$ 
with a five-loop beta  function~\cite{Kleinert}:
\bqa
\label{rg2}
\mu{\partial\over \partial\mu}\alpha=
3\alpha^2-{17\over3}\alpha^3+32.54\alpha^4-271.6\alpha^5+
2848.6\alpha^6\;.
\eqa
The lack of convergence of the weak-coupling expansion for large
coupling
is evident in Fig.~\ref{fpert}.
The band obtained by varying $\mu$ by a factor of two is not 
necessarily a good measure of the error, 
but it is certainly a lower bound on the theoretical error.
Another indicator of the  theoretical error is the deviation 
between successive approximations.  We can infer from Fig.~\ref{fpert} 
that the error grows rapidly when $g(2 \pi T)$ exceeds 1.5.

\begin{figure}[t]
\epsfysize=8cm
\centerline{\epsffile{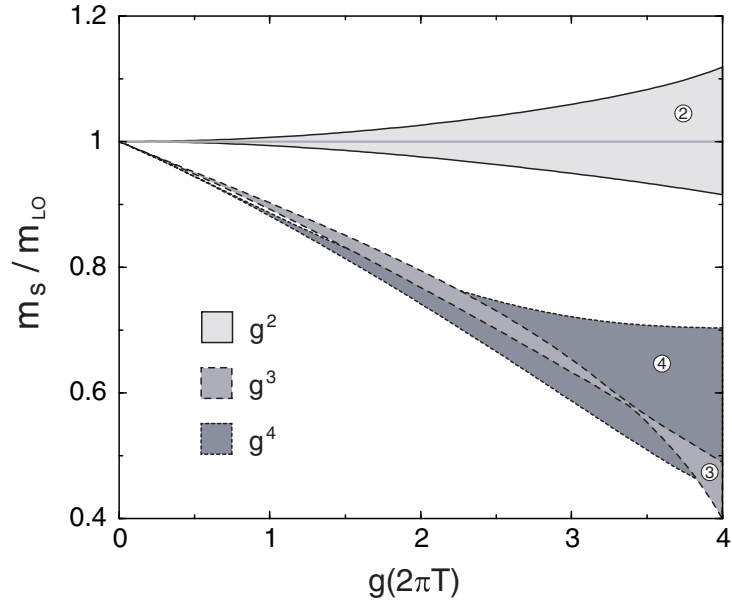}}
\vspace{3mm}
\caption[a]{Weak-coupling expansion to orders $g^2$, $g^3$, and $g^4$
for the screening mass normalized to the leading-order expression as a
function of $g(2\pi T)$.}
\label{mspert}
\end{figure}

A similar behavior can be seen in the weak-coupling expansion for the 
screening mass, which has been calculated to order 
$g^4$~\cite{Braaten-Nieto:scalar}.
In Fig.~\ref{mspert}, we show the screening mass $m_s$ 
normalized to the leading order result $m_{\rm LO}=g(2\pi T)T/\sqrt{24}$
as a function of $g(2\pi T)$, for each of the three
successive approximations to $m_s^2$.
The bands correspond to varying $\mu$
from $\pi T$ to $4\pi T$. The poor convergence is again evident.
The pattern is similar to that in Fig.~\ref{fpert}, 
with a large deviation between the order-$g^2$ and order-$g^3$ 
approximations and a large increase in the size of the band for $g^4$.

There are several ways to reorganize perturbation theory to improve its
convergence. One method is Pad\'e approximants~\cite{pade}.
This method is limited 
to observables like the pressure,
for which several terms in the weak-coupling expansion are known.
Its application is further complicated by the appearance of logarithms
of the coupling constant in the coefficients of the weak-coupling expansion.
However, the greatest problem with Pad\'e approximants is that, with
no understanding of the analytic behavior of ${\cal P}$ at strong coupling,
it is little more than a numerological recipe.

An alternative with greater physical motivation 
is a self-consistent approach \cite{Baym}.
In this method, perturbation theory is reorganized by expressing the free energy 
as a stationary point of a functional $\Omega$ 
of the exact self-energy function $\Pi(p_0,{\bf p})$ 
called the thermodynamic potential \cite{Luttinger-Ward}.
Since the exact self-energy is not known, $\Pi$ can be regarded as a 
variational function. 
The ``$\Phi$-derivable'' prescription of Baym~\cite{Baym}
is to truncate the perturbative expansion 
for the thermodynamic potential $\Omega$ and to determine
$\Pi$ self-consistently as a stationary point of $\Omega$. This gives an
integral equation for the self-energy that is hard to solve numerically,
unless $\Pi$ is momentum independent. 
A more tractable approach is to find an approximate solution
to the integral equations that is accurate only in the weak-coupling limit.
Such an approach has been applied by Blaizot, Iancu, and
Rebhan to massless scalar field theories and gauge 
theories~\cite{comp1,BIR}.

Another variational approach is 
{\it screened perturbation theory} (SPT) introduced by Karsch, Patk\'os
and Petreczky~\cite{K-P-P}. This approach is less ambitious than the 
$\Phi$-derivable approach.
Instead of introducing a variational function, it introduces a single
variational parameter $m$. This parameter has a simple and obvious
physical interpretation as a thermal mass. 
The advantage of screened perturbation theory is that it is 
straightforward to apply.
Higher order corrections are calculable, so one can test whether it improves
the convergence of the weak-coupling expansion.
Karsch, Patk\'os and Petreczky applied screened perturbation theory
to a massless scalar field theory with a $\phi^4$ interaction,
computing the two-loop
pressure and the three-loop pressure in the large-$N$ limit.


In Ref.~\cite{spt}, a detailed study of screened perturbation theory
for a massless scalar field theory was presented. The pressure and entropy
were calculated to three loops and the screening mass to two loops.
It was shown that, in contrast to the weak-coupling expansions, the 
SPT-improved approximations converge even for rather large values of the
coupling constant. 
In Ref.~\cite{spt}, the sum-integrals for the three-loop free energy were 
evaluated exactly by replacing the sums by contour integrals,
extracting the poles in $\epsilon$, and then reducing the momentum 
integrals to integrals that were at most three-dimensional 
and could be evaluated numerically.   The resulting expressions,
while truncated in the coupling constant were ``exact'' in the sense
that they included contributions from all orders in $m/T$.
In this paper we continue the study of screened perturbation theory by
performing an analytic expansion
of the sum-integrals in powers of $m/T$ and demonstrate
that the first few terms in the expansion 
give an accurate approximation to the exact SPT result.

The paper is organized as follows. In section II, 
we describe the systematics of screened perturbation theory.
In section III, we calculate the free energy and entropy to three loops,
and the screening mass to two loops,
expanding in powers of $m/T$. In Section IV, we 
calculate the screening mass to two loops using the $m/T$ expansion.
In Section V, we briefly discuss the two-loop tadpole gap that 
generalizes the one-loop gap equation.
In Section VI, we 
study the convergence properties of SPT-improved 
results for the pressure, entropy, and screening mass
using the $m/T$ expansion.
Finally, in section VII, we summarize and conclude.
Necessary calculational details are collected in
four appendices.

\section{Screened Perturbation Theory}
The Lagrangian density for a massless scalar field with a $\phi^4$ interaction is
\bqa
\label{ori}
{\cal L}={1\over2}\partial_{\mu}\phi\partial^{\mu}\phi
-{g^2\over24}\phi^4+\Delta{\cal L}\;,
\eqa
where $g$ is the coupling constant and $\Delta{\cal L}$
includes counterterms.
Renormalizability guarantees that $\Delta{\cal L}$ is of the form
\bqa
\Delta{\cal L} = {1\over2} \Delta Z \, \partial_{\mu}\phi\partial^{\mu}\phi
	- {1\over24} \Delta g^2 \phi^4 \; .
\eqa

Screened perturbation theory, which was introduced by Karsch, Patk\'os
and Petreczky~\cite{K-P-P}, is simply a reorganization of the perturbation
series for thermal field theory.
It can be made more systematic by using a framework called 
``optimized perturbation theory'' that Chiku and Hatsuda~\cite{Chiku-Hatsuda}
have applied to a spontaneously broken scalar field theory. 
The Lagrangian density is written as 
\bqa
\label{SPT}
{\cal L}_{\rm SPT}=-{\cal E}_0+{1\over2}\partial_{\mu}\phi\partial^{\mu}\phi
-{1\over2}\left(m^2-m_1^2\right)\phi^2
-{g^2\over24}\phi^4
+\Delta{\cal L}+\Delta{\cal L}_{\rm SPT}
\;.
\eqa
Here, ${\cal E}_0$ is the vacuum energy density term, and we have added and
subtracted mass terms. If we set ${\cal E}_0=0$ and $m_1^2=m^2$, we recover
the original Lagrangian Eq.~(\ref{ori}).
Screened perturbation theory is defined by taking $m^2$ to be of order unity
and $m_1^2$ to of order $g^2$, expanding systematically in powers of
$g^2$ and setting $m_1^2=m^2$ at the end of the calculation.
This defines a reorganization of the perturbative series in which the 
expansion is about the free field theory defined by
\bqa
{\cal L}_{\rm free}=
-{\cal E}_0+{1\over2}\partial_{\mu}\phi\partial^{\mu}\phi
-{1\over2}m^2\phi^2\;.
\eqa
The interacting term is 
\bqa
{\cal L}_{\rm int}=
{1\over2}m_1^2\phi^2
-{g^2\over24}\phi^4
+\Delta{\cal L}+\Delta{\cal L}_{\rm SPT}\;.
\eqa
Screened perturbation theory generates new ultraviolet divergences, but they
can be cancelled by the additional counterterm in $\Delta{\cal L}_{\rm SPT}$.
If we use dimensional regularization and minimal subtraction, the coefficients
of these operators are polynomials in $g^2$ and $(m^2-m_1^2)$.
The additional counterterms required to remove the new divergences are
\bqa
\Delta{\cal L}_{\rm SPT}=-\Delta{\cal E}_0-{1\over2}\left(
\Delta m^2-\Delta m_1^2
\right)\phi^2\;.
\eqa
Several terms in the power series expansions of the
counterterms are known from previous calculations at zero temperature.
The counterterms $\Delta g^2$ and $\Delta m^2$ are known to order 
$\alpha^5$~\cite{Kleinert}. 
We will need the coupling constant counterterm only to leading
order in $\alpha$:
\bqa
\Delta g^2=\left[
{3\over2\epsilon}\alpha+...
\right]g^2\;.
\eqa
We need the mass counterterms $\Delta m^2$ and $\Delta m^2_1$
to next-to-leading order and 
leading order in $\alpha$, respectively:
\bqa
\label{dmm}
\Delta m^2&=&\left[{1\over2\epsilon}\alpha+\left({1\over2\epsilon^2}
-{5\over24\epsilon}\right)\alpha^2+...
\right]m^2\;, \\
\label{d1m1}
\Delta m^2_1&=&\left[{1\over2\epsilon}\alpha+...
\right]m^2_1\;.
\eqa
The counterterm for $\Delta{\cal E}_0$
has been calculated to order $\alpha^4$~\cite{kastening}.
We will need its expansion only to second order 
in $\alpha$ and $m_1^2$:
\bqa\nonumber
(4\pi)^2\Delta{\cal E}_0&=&\left[
{1\over4\epsilon}+{1\over8\epsilon^2}\alpha
+\left(
{5\over48\epsilon^3}-{5\over72\epsilon^2}+{1\over96\epsilon}
\right)\alpha^2
\right]m^4
\\&&
\label{de}
-2\left[{1\over4\epsilon}+{1\over8\epsilon^2}\alpha\right]m_1^2m^2
+{1\over4\epsilon}m_1^4\;.
\eqa

\section{Free energy to three loops}

\begin{figure}[ht]
\epsfysize=3cm
\centerline{\epsffile{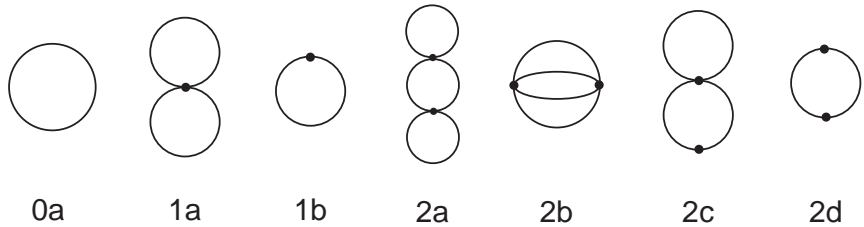}}
\vspace{3mm}
\caption[a]{Diagrams for the one-loop (0a), two-loop (1a and 1b), and
three-loop (2a, 2b, 2c, and 2d) free energy.}
\label{diagrams}
\end{figure}

In this section, we calculate the $m/T$ expansions of the 
pressure and entropy density
to three loops in screened perturbation theory.  In performing
the truncation $m$ is treated as a quantity that is ${\cal O}(g)$
and include all terms which contribute to order $g^5$.
The diagrams for the free energy that are included at this order 
are those shown in Fig.~\ref{diagrams} 
together with diagrams involving counterterms.

\subsection{One-loop free energy}
The free energy at leading order in $g^2$ is
\bqa
{\cal F}_0&=&{\cal E}_0+{\cal F}_{\rm 0a}+\Delta_0{\cal E}_0\;,
\eqa
where $\Delta_0{\cal E}_0$ is the term of order $g^0$ in the vacuum energy
counterterm Eq.~(\ref{de}). The expression for diagram 0a in Fig.~\ref{diagrams} is 
\bqa
{\cal F}_{\rm 0a}&=&{1\over2}\sumint_{P}\log\left[P^2+m^2\right]\;,
\eqa
with $\sumintsmall_P$ defined in Appendix \ref{appa}.

Treating $m$ as ${\cal O}(g)$ and including all terms which contribute
through ${\cal O}(g^5)$, we obtain
\bqa
{\cal F}_{\rm 0a}&=&
{1\over2}{\cal I}_0^{\prime}
+{1\over2}m^2{\cal I}_1
+{1\over2}TI_0^{\prime}
-{1\over4}m^4{\cal I}_2
+{\cal O}(m^6/T^2)\;,
\eqa
where ${\cal I}^\prime_0$ and ${\cal I}_n$ are defined in
Appendix \ref{appa} and $I^\prime_0$ is defined in Appendix
\ref{appb}.
 
The resulting expression is logarithmically divergent and the 
pole in $\epsilon$ is cancelled by the zeroth order term 
$\Delta_0{\cal E}_0$ in Eq.~(\ref{de}).
The final result for the truncated one-loop free energy is
\bqa
{\cal F}_0&=&
-{\pi^2T^4\over90}
\Bigg[1-15\hat{m}^2+60\hat{m}^3+45\hat{m}^4(L+\gamma)\Bigg]\;,
\label{f0}
\eqa
where $\hat{m}={m\over2\pi T}$ and $L=\log{\mu\over4\pi T}$.

\subsection{Two-loop free energy}
The contribution to the free energy of order $g^2$ is
\bqa
\label{f1}
{\cal F}_1&=&{\cal F}_{\rm 1a}
+{\cal F}_{\rm 1b}
+\Delta_1{\cal E}_0
+{\partial{\cal F}_{\rm 0a}\over\partial m^2}\Delta_1m^2\;,
\eqa
where $\Delta_1{\cal E}_0$ and $\Delta_1m^2$ are the counterterms
of order $g^2$, respectively.
The expressions for the diagrams 1a and 1b in Fig.~\ref{diagrams} are
\bqa
\label{1a0}
{\cal F}_{\rm 1a}&=&{1\over8}g^2\left(\sumint_{P}{1\over P^2+m^2}\right)^2\;,
\\
\label{1b0}
{\cal F}_{\rm 1b}&=&-{1\over2}m_1^2\sumint_{P}{1\over P^2+m^2}\;.
\eqa
The sum-integrals in Eqs.~(\ref{1a0}) and~(\ref{1b0}) are expanded to the
order required:
\bqa
\label{f1a}
{\cal F}_{\rm 1a}&=&{1\over8}g^2
\Bigg[
{\cal I}_1^2
+2T{\cal I}_1I_1
-2m^2{\cal I}_1{\cal I}_2+T^2I_1^2
-2m^2T{\cal I}_2I_1
\Bigg]\;,
\\
\label{f1b}
{\cal F}_{\rm 1b}&=&
-{1\over2}m_1^2\Bigg[
{\cal I}_1+TI_1-m^2{\cal I}_2
\Bigg]\;,
\eqa
where $I_n$ is defined in Appendix \ref{appb}.

The poles in $\epsilon$ in Eqs.~(\ref{f1a}) 
and~(\ref{f1b}) are cancelled by the counterterms
in Eq.~(\ref{f1}). 
The final result for the truncated two-loop free energy is
\bqa\nonumber
{\cal F}_1&=&
{g^2T^4\over1152}
\Bigg[
1-12\hat{m}-12\hat{m}^2\left(
L+\gamma-3\right)
+72(L+\gamma)\hat{m}^3
\Bigg]\\ 
&&-{m_1^2T^2\over24}
\Bigg[1-6\hat{m}-6\hat{m}^2(L+\gamma)\Bigg]\;.
\label{2free}
\eqa
\subsection{Three-loop free energy}
The contribution to the free energy of order $g^4$ is 
\bqa\nonumber
{\cal F}_2&=&{\cal F}_{\rm 2a}+{\cal F}_{\rm 2b}
+{\cal F}_{\rm 2c}+{\cal F}_{\rm 2d}+\Delta_2{\cal E}_0
+{\partial{\cal F}_{\rm 0a}\over\partial m^2}\Delta_2m^2
+{1\over2}{\partial^2{\cal F}_{\rm 0a}\over\left(\partial m^2\right)^2}
\left(\Delta_1m^2\right)^2
\\ &&
\label{f2}
+\left({\partial{\cal F}_{\rm 1a}\over\partial m^2}
+{\partial{\cal F}_{\rm 1b}\over\partial m^2}
\right)\Delta_1m^2
+{{\cal F}_{\rm 1a}\over g^2}\Delta_1g^2
+{{\cal F}_{\rm 1b}\over m_1^2}\Delta_1m_1^2\;,
\eqa
where we have included all necessary counterterms.
The expressions for the diagrams $2a$, $2b$, $2c$, and $2d$ 
in Fig.~\ref{diagrams} are
\bqa
{\cal F}_{\rm 2a}&=&
-{1\over16}g^4\left(\sumint_{P}{1\over P^2+m^2}\right)^2
\sumint_{Q}{1\over(Q^2+m^2)^2}\;, \\
{\cal F}_{\rm 2b}&=&
-{1\over48}g^4\sumint_{PQR}
{1\over P^2+m^2}
{1\over Q^2+m^2}{1\over R^2+m^2}{1\over (P+Q+R)^2+m^2}
\;,\\
{\cal F}_{\rm 2c}&=& 
{1\over4}g^2m_1^2\sumint_{P}{1\over P^2+m^2}
\sumint_{Q}{1\over(Q^2+m^2)^2}
\;,\\
{\cal F}_{\rm 2d}&=&
-{1\over4}m_1^4\sumint_{P}{1\over(P^2+m^2)^2}
\;.
\eqa
Expanding in powers of $m^2$ to the appropriate order gives
\bqa
{\cal F}_{\rm 2a}&=& 
-{1\over16}g^4\Bigg[
T{\cal I}_1^2I_2
+{\cal I}_1^2{\cal I}_2+2T^2{\cal I}_1I_1I_2
+T^3I_1^2I_2
+2TI_1{\cal I}_1{\cal I}_2
\label{2a}
-2m^2T{\cal I}_1{\cal I}_2I_2
\Bigg]\;,\\
\label{2b}
{\cal F}_{\rm 2b}&=& 
-{1\over48}g^4\Bigg[{\cal I}_{\rm ball}+
T^3I_{\rm ball}+4T I_1 {\cal I}_{\rm sun}(0)
\Bigg]
\;,\\
\label{2c}
{\cal F}_{\rm 2c}&=& 
{1\over4}g^2m_1^2\Bigg[
T{\cal I}_1I_2+{\cal I}_1{\cal I}_2
+T^2I_1I_2+T{\cal I}_2I_1-m^2T{\cal I}_2I_2
\Bigg]
\;,\\
\label{2d}
{\cal F}_{\rm 2d}&=& 
-{1\over4}m_1^4\Bigg[
{\cal I}_2+TI_2
\Bigg]\;,
\eqa
where $I_{\rm ball}$, ${\cal I}_{\rm sun}$, and ${\cal I}_{\rm ball}$
are defined in Appendices \ref{appb}, \ref{appc}, \ref{appd} respectively.

The poles in $\epsilon$ in Eqs.~(\ref{2a})-(\ref{2d}) 
are cancelled by the counterterms
in Eq.~(\ref{f2}). 
The final result for the truncated three-loop free energy is

\bqa\nonumber
{\cal F}_2&=&
-{g^4T^4\over2304(4\pi)^2\hat{m}}\Bigg[
1-2\hat{m}
\Bigg(
{59\over15}-\gamma-3L+2{\zeta^{\prime}(-3)\over\zeta(-3)}
-4{\zeta^{\prime}(-1)\over\zeta(-1)}
\Bigg)
\\\nonumber
&& 
\hspace{5mm}
-12\hat{m}^2
\Bigg(
5+7L+3\gamma
-8\log\hat{m}-8\log2
-4{\zeta^{\prime}(-1)\over\zeta(-1)}
\Bigg)
\Bigg]
\\\nonumber
&&
\hspace{5mm}
+{g^2m_1^2T^2\over48(4\pi)^2\hat{m}}\Bigg[
1+2\hat{m}\left(L+\gamma-3\right)
-18\hat{m}^2(L+\gamma)
\Bigg]
\\&&
\hspace{5mm}
-{m_1^4\over64\hat{m}}\Bigg[1+2\hat{m}(L+\gamma)\Bigg]
\;.
\label{3free}
\eqa

\subsection{Pressure to three loops}
The pressure ${\cal P}$ is given by $-{\cal F}$. 
The contributions to the pressure of zeroth, first, and second order in
$g^2$ are given by Eqs.~(\ref{f0}),~(\ref{2free}), and~(\ref{3free}), 
respectively. Adding
these contributions and setting ${\cal E}_0=0$ and $m_1^2=m^2$, we obtain 
approximations to the pressure in screened perturbation theory which are 
accurate to ${\cal O}(g^5)$.

\noindent
The one-loop approximation to the pressure is
\bqa
{\cal P}_{0} &=&
{\cal P}_{\rm ideal} \Bigg[
1-15\hat{m}^2+60\hat{m}^3+45\hat{m}^4(L+\gamma)
\Bigg]
\;.
\label{1p}
\eqa

\noindent
The two-loop approximation to the pressure is obtained by adding 
Eq.~(\ref{2free}) with $m_1^2=m^2$:
\bqa\nonumber
{\cal P}_{0+1}&=&
{\cal P}_{\rm ideal} \Bigg\{
1-{5\over4}\alpha
+15\hat{m}\alpha
+15\hat{m}^2(L+\gamma-3)\alpha
\\&&
\hspace{2cm}
-30\hat{m}^3\bigg[1+3(L+\gamma)\alpha\bigg]
-45\hat{m}^4(L+\gamma)\Bigg\}
\;.
\label{2p}
\eqa
   
\noindent
The three-loop approximation to the pressure is obtained by adding
Eq.~(\ref{3free}) with $m_1^2=m^2$:
\bqa\nonumber
{\cal P}_{0+1+2}&=&
{\cal P}_{\rm ideal} \Bigg\{
1-{5\over4}\alpha
+\Bigg[
-{59\over12}+{15\over4}L+{5\over4}\gamma
-{5\over2}{\zeta^{\prime}(-3)\over\zeta(-3)}
+5{\zeta^{\prime}(-1)\over\zeta(-1)}
\Bigg]\alpha^2
\\\nonumber
&& \hspace{5mm}
+{15\over2}\hat{m}\Bigg[
1-\Bigg(
5+3\gamma+7L
-8\log\hat{m}
-8\log2
-4{\zeta^{\prime}(-1)\over\zeta(-1)}
\Bigg)\alpha
\Bigg]\alpha
\\ 
&& \hspace{1cm}
-{15\over2}\hat{m}^3\Bigg[1-6(L+\gamma)\alpha\Bigg]
+{5\over8\hat{m}}\alpha^2 \Bigg\}
\label{3p}
\;.
\eqa
Note that if we substitute the leading-order result for the screening mass, 
$m=g(2\pi T)T/\sqrt{24}$, in Eq.~(\ref{3p}), we recover the 
weak-coupling expansion through order $g^5$.

\subsection{Entropy to three loops}
Given a diagrammatic expansion for the free energy ${\cal F}$, the entropy
density ${\cal S}$ has a diagrammatic expansion defined by
\bqa
S_{\rm diag}&=&
-{\!\!\partial\over\partial T}{\cal F}(T,g,m,m_1,\mu)\;,
\eqa
where the partial derivative is taken with all the other variables
$g$, $m$, $m_1$, and $\mu$ held fixed.
The one-, two-, and three-loop approximations to ${\cal S}$ are then obtained
by taking partial derivatives of the corresponding expressions for the 
pressure ${\cal P}$.

\noindent
The one-loop approximation to the entropy ${\cal S}$ is 
obtained by differentiating Eq.(\ref{1p})
\bqa
\label{s0}
{\cal S}_0 &=&
{\cal S}_{\rm ideal} \Bigg[
1-{15\over2}\hat{m}^2
+15\hat{m}^3
-{45\over4}\hat{m}^4 \Bigg]
\;,
\eqa
where ${\cal S}_{\rm ideal}=(2\pi^2/45)T^3$.

\noindent
The two-loop approximation to the the entropy ${\cal S}$ is
obtained by differentiating Eq.(\ref{2p})
\bqa\nonumber
{\cal S}_{0+1}&=&
{\cal S}_{\rm ideal} \Bigg\{
1-{5\over4}\alpha
+{45\over4}\hat{m}\alpha
+{15\over2}\hat{m}^2\left(L+\gamma-{7\over2}\right)\alpha
\\&&
\hspace{2cm}
-{15\over2}\hat{m}^3\bigg[1+3\left(L+\gamma-1\right)\alpha\bigg]
+{45\over4}\hat{m}^4 \Bigg\}
\;.
\label{s1}
\eqa

\noindent
The three-loop approximation to the the entropy ${\cal S}$ is
obtained by differentiating Eq.(\ref{3p})
\bqa\nonumber
{\cal S}_{0+1+2}&=&
{\cal S}_{\rm ideal} \Bigg\{
1-{5\over4}\alpha
+\Bigg(
-{281\over48}+{5\over4}\gamma+{15\over4}L+5
{\zeta^{\prime}(-1)\over\zeta(-1)}
-{5\over2}{\zeta^{\prime}(-3)\over\zeta(-3)}
\Bigg)\alpha^2
\\\nonumber
&& \hspace{5mm}
+{45\over8}\hat{m}\Bigg[
1-\Bigg(
{16\over3}+7L+3\gamma-8\log\hat{m}-
4{\zeta^{\prime}(-1)\over\zeta(-1)}
-8\log2
\Bigg)\alpha
\Bigg]\alpha
\\ 
&& \hspace{1cm}
-{15\over8}\hat{m}^3\Bigg[1-6(L+\gamma-1)\alpha\Bigg]
\label{3pp}
+{25\over32\hat{m}} \alpha^2 \Bigg\}
\;.
\label{s2}
\eqa

\section{Screening mass to two loops}
\begin{figure}[ht]
\epsfysize=2.5cm
\centerline{\epsffile{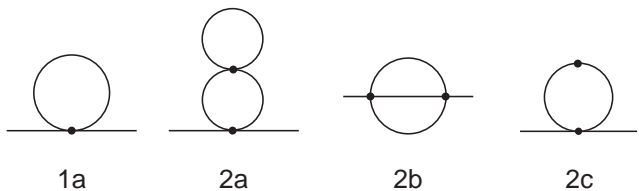}}
\vspace{3mm}
\caption[a]{Diagrams for the one-loop (1a and 1b) and two-loop (2a, 
2b and 2c) self-energy.}
\label{selfdiagrams}
\end{figure}

In this section, we calculate the $m/T$ expansion of the screening mass to two loops. 
The diagrams for the self energy that are included at this order 
are those shown in Fig.~\ref{selfdiagrams} 
together with diagrams involving counterterms. 

The screening mass $m_s$ is defined by the location of the pole of the static propagator:
\bqa
\label{scrdef}
{\bf p}^2+m^2+\Pi(0,{\bf p})=0\hspace{2cm} \mbox{at}\hspace{1cm} 
{\bf p}^2=-m^2_s\;,
\eqa
where $\Pi(p_0,{\bf p})$ is the self-energy function.
This equation can be solved order-by-order in powers of $\alpha$ and $m_1^2$.
The solution at zeroth order in $g^2$ is simply $m_s^2=m^2$.

\subsection{One-loop self-energy}
The self-energy to leading order in $g^2$ is
\bqa
\Pi_1=\Pi_{\rm 1a}-m_1^2+\Delta_1m^2\;,
\eqa
where $\Delta_1m^2$ is the mass counterterm of order $\alpha$ given 
in Eq.~(\ref{dmm}). 
The expression for the diagram 1a in Fig.~\ref{selfdiagrams} is
\bqa\nonumber
\Pi_{\rm 1a}
&=&{1\over2}g^2\sumint_P{1\over P^2+m^2}\;.
\eqa
This diagram is expanded to second order in $m^2$:
\bqa
\label{1a}
\Pi_{\rm 1a}&=&
{1\over2}g^2\Bigg[
{\cal I}_1+TI_1-m^2{\cal I}_2
\Bigg]\;.
\eqa
The pole in Eq.~(\ref{1a}) is cancelled by the counterterm $\Delta_1m^2$.
The final result for the one-loop self-energy is
\bqa
\label{pi1}
\Pi_{\rm 1}=
{g^2T^2\over24}
\Bigg[
1-6\hat{m}
-6\hat{m}^2\left(L+\gamma\right)
\Bigg]
-m_1^2\;.
\eqa

\subsection{Two-loop self-energy}

The contribution to the self-energy of second order in $g^2$ is
\bqa
\label{2scree}
\Pi_{2}(P)=\Pi_{\rm 2a}+\Pi_{\rm 2b}(P)+\Pi_{\rm 2c}
+{\partial\Pi_{\rm 1a}\over\partial m^2}\Delta_1m^2
+{\Pi_{\rm 1a}\over g^2}\Delta_1g^2+\Delta_2m^2
-\Delta_1m_1^2\;.
\eqa
The expressions for the diagrams 2a, 2b, and 2c in Fig.~\ref{selfdiagrams} are
\bqa
\label{2la}
\Pi_{\rm 2a}&=&-{1\over4}g^4\sumint_{Q}{1\over Q^2+m^2}
\sumint_R{1\over(R^2+m^2)^2}\;,\\
\label{2lb}
\Pi_{\rm 2b}(P)&=&-{1\over6}g^4\sumint_{QR}{1\over Q^2+m^2}{1\over R^2+m^2 }{1\over (P+Q+R)^2+m^2}\;,\\
\label{2lc}
\Pi_{\rm 2c}&=&{1\over2}g^2m_1^2\sumint_{Q}{1\over(Q^2+m^2)^2}\;.
\eqa
The diagrams $\Pi_{\rm 2a}$ and $\Pi_{\rm 2c}$ are independent of the momentum $P$. 
%
Expanding to first order in $m^2$, we obtain
\bqa
\label{2ade}
\Pi_{\rm 2a}&=&-{1\over4}g^4
\Bigg[
T{\cal I}_1I_2+{\cal I}_1{\cal I}_2+T^2I_1I_2
\Bigg]
\;. \\
\Pi_{\rm 2c}&=&
{1\over2}g^2m_1^2
\Bigg[
{\cal I}_2+TI_2
\Bigg]
\eqa
The diagram $\Pi_{\rm 2b}$ depends on the external momentum $P$.
The equation~(\ref{scrdef}) for the screening mass involves the self-energy
at $p_0=0$. To calculate the screening mass to second order in $g^2$, 
we need the analytic  continuation 
of $\Pi(0,{\bf p})$ to ${\bf p}^2=-m^2$.
The diagram is calculated in Appendix \ref{appc}. The result is
\bqa
\Pi_{\rm 2b}(0,{\bf p})\Bigg|_{{\bf p}^2=-m^2}&=&
-{1\over6}g^4\Bigg[{\cal I}_{\rm sun}(0)
+I_{\rm sun}
\Bigg]\;,
\label{2bde}
\eqa
where $I_{\rm sun}$ is defined in Appendix \ref{appb}.

The poles in Eqs.~(\ref{2ade})-(\ref{2bde}) are cancelled by the counterterms
in Eq.~(\ref{2scree}). The final result for the truncated two-loop 
contribution to the self-energy at $p_0=0$ and ${\bf p}^2=-m^2$ is
\bqa\nonumber
\Pi_2(0,{\bf p})\Bigg|_{{\bf p}^2=-m^2}
&=&-{g^4T^2\over768\pi^2\hat{m}}
\Bigg\{1+
2\hat{m}\left[3L+\gamma+1-4\log\hat{m}-8\log2-
2{\zeta^{\prime}(-1)\over\zeta(-1)}\right] 
\Bigg\}\\
&&
\label{pi2}
\hspace{2cm}+
{g^2m_1^2\over32\pi^2\hat{m}}
\Bigg[1+2\left(L+\gamma\right)\hat{m}
\Bigg]\;.
\eqa

\subsection{Screening mass}
Since the dependence of the self-energy on the momentum enters only at
order $g^4$ and since the leading-order solution to the screening mass is 
$m_s=m$, the solution to the equation~(\ref{scrdef}) 
to order $g^4$ is simply

\bqa
\label{screexpl}
m_s^2=m^2+\Pi(0,{\bf p}^2)\Bigg|_{{\bf p}^2=-m^2}\;.
\eqa
The result for the one-loop screening mass is
\bqa
\label{onesc}
\hat{m}_s^2=
{1\over6}\alpha
\Bigg[1- 6\hat{m}
-6\hat{m}^2\left(
L+\gamma\right)
\Bigg]\;.
\eqa
%

The solution to order $g^4$ is obtained by inserting the sum of 
Eqs.~(\ref{pi1}) and~(\ref{pi2}) into Eq~(\ref{screexpl}). Setting
$m_1^2=m^2$, the result is
\bqa
\hat{m}_s^2&=&
{1\over6}\alpha
\Bigg\{1
-{1\over2\hat{m}}\alpha
-\left[3L+\gamma+1+4\log\hat{m}-8\log2-
2{\zeta^{\prime}(-1)\over\zeta(-1)}\right]\alpha - 3\hat{m}\Bigg\}\;.
\label{scr2}
\eqa

If we substitute the leading-order result for the screening mass, 
$m=g(2\pi T)T/\sqrt{24}$, in Eq.~(\ref{scr2}), we recover the 
weak-coupling expansion through order $g^4$~\cite{Braaten-Nieto:scalar}.

\section{Gap Equation}\label{tadders}
The mass parameter $m$ in screened perturbation theory is completely 
arbitrary.
To complete the calculation it is necessary
to specify $m$ as a function of $g$ and $T$. 
One of the complications from the ultraviolet divergences is that
the parameters ${\cal E}_0$, $m^2$, $g^2$, and $m_1^2$ all become
running parameters that depend on a renormalization scale $\mu$.
In our prescription for recovering the original theory, we must therefore
specify the renormalization scale $\mu_*$
at which the Lagrangian~(\ref{SPT}) reduces to Eq.~(\ref{ori}).
The prescription can
be written
\bqa
{\cal E}_0(\mu_*)&=0& \, , \\ 
m^2(\mu_*)&=&m_1^2(\mu_*)=m^2_{*}(T),
\eqa
where $m_{*}(T)$ is some prescribed function of the temperature.

\begin{figure}[t]
\epsfxsize=15.5cm
\centerline{\epsffile{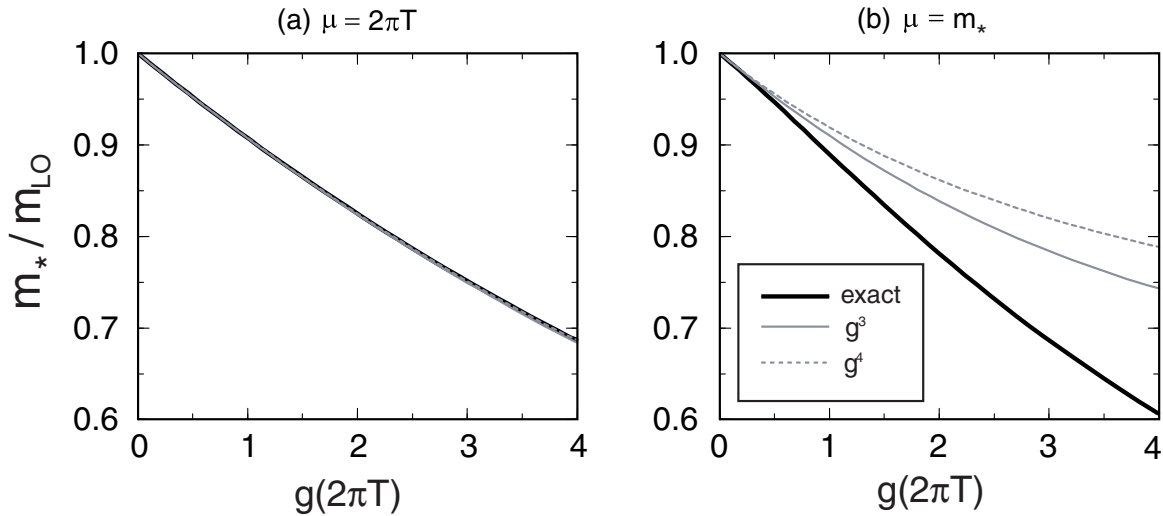}}
\vspace{3mm}
\caption[a]{Solutions $m_*(T)$ to the one-loop tadpole
gap equation as a function of
$g(2\pi T)$ for (a) $\mu=2\pi T$ and (b) $\mu=m_*$.  
Exact SPT curves are taken from Ref.~\cite{spt}.
}
\label{tadgap}
\end{figure}

The prescription of
Karsch, Patk\'os, and Petreczky for $m_*(T)$ is the solution to the
one-loop gap equation:
\bqa
\label{pet}
m_{*}^2={1\over2}\alpha(\mu_*)\left[
J_1(m_*/T)T^2-\left(2\log{\mu_*\over m_*}+1\right)m_*^2
\right]
\;.
\eqa
Their choice for the scale was $\mu_*=T$.
The function $J_1(\beta m)$ is defined as
\bqa
J_1(\beta m)&=&8\beta^{2}
\int_0^{\infty}dk\;{k^{2}\over\left(k^2+m^2\right)^{1/2}}
{1\over e^{\beta\left(k^2+m^2\right)^{1/2}}-1}\;. 
\eqa
In the limit $\beta m\longrightarrow 0$, this
integral reduces to
\bqa
J_1(\beta m) &\longrightarrow& {4\pi^2\over3}
	\;-\; 4\pi \beta m
	\;-\; 2 \left( \log{\beta m\over4\pi}-{1\over2}+\gamma\right) 
		(\beta m)^2 \;.
\eqa
In the same limit, 
Eq.~(\ref{pet}) reduces to
\bqa
\hat{m}_*^2={1\over6}\alpha\Bigg[1-6\hat{m}_*
-6\hat{m}^2_*\left(L+\gamma\right)\Bigg]\;.
\label{oneloopgapeq}
\eqa
The one-loop gap equation is identical 
to the one-loop screening mass if we choose $m=m_s=m_*$

Various mass prescriptions that generalize 
Eq.~(\ref{pet})
were extensively studied in Ref.~\cite{spt}.
In this paper, we confine ourselves to using the tadpole mass which is
defined by $m_t^2={1\over2}g^2\langle\phi^2\rangle$. This can also be expressed as
a derivative of the free energy:
\bqa
\label{tad}
m_t^2=g^2{\!\!\!\!\!\!\partial\over\partial m^2}{\cal F}(T,g,m,m_1,\mu)\Bigg|_{m_1=m}\;,
\eqa
where the partial derivative is taken before setting $m_1=m$.

The one-loop expression for the tadpole mass is given differentiating
Eq.~(\ref{f0}):
\bqa
\label{1g}
\hat{m}_t^2={1\over6}\alpha\Bigg[1-6\hat{m}
-6\hat{m}^2\left(L+\gamma\right)
\Bigg]
\;.
\eqa

\begin{figure}[ht]
\epsfxsize=15.5cm
\centerline{\epsffile{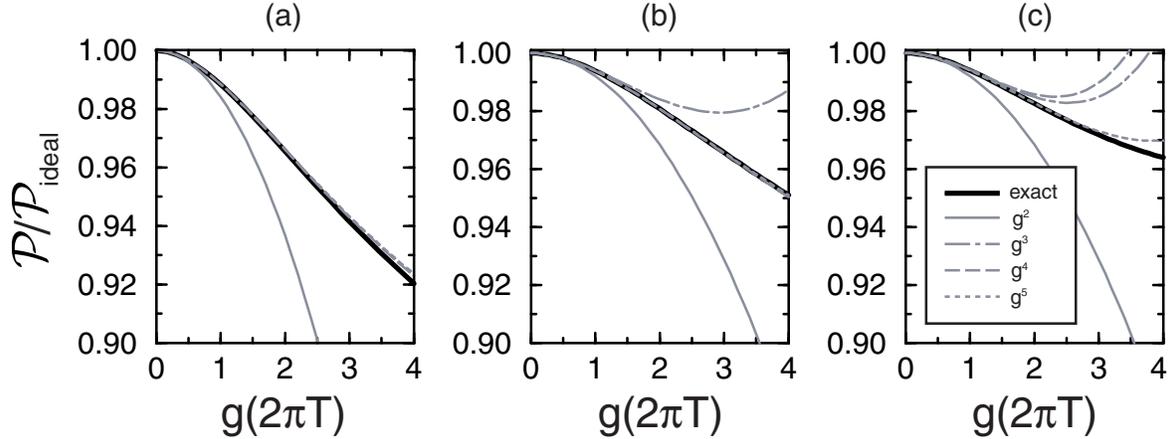}}
\vspace{3mm}
\caption[a]{The one-, two and three-loop SPT improved approximations to the
pressure as a function of $g(2\pi T)$ for $\mu=2\pi T$.
Exact SPT curves are taken from Ref.~\cite{spt}.}
\label{p0+1}
\end{figure}

\begin{figure}[ht]
\epsfxsize=15.5cm
\centerline{\epsffile{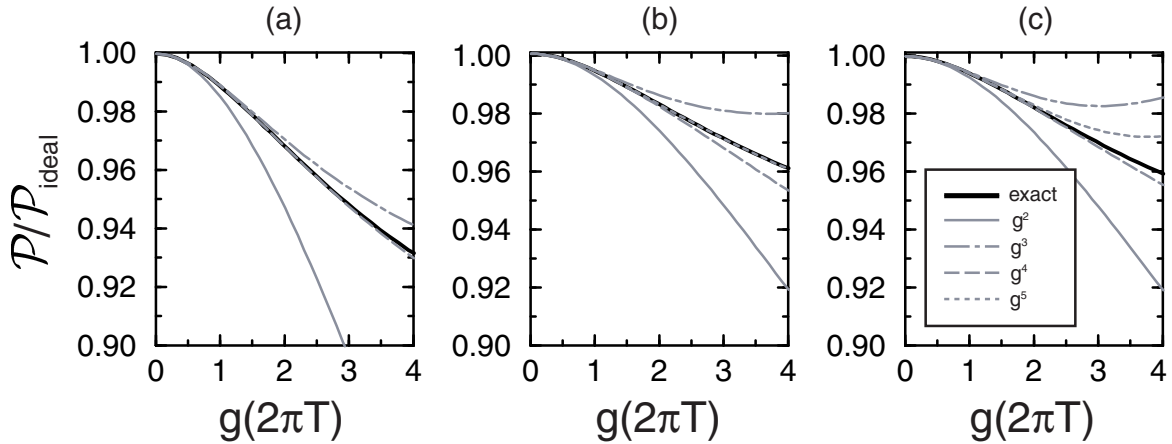}}
\vspace{3mm}
\caption[a]{The one-, two and three-loop SPT improved approximations to the
pressure as a function of $g(2\pi T)$ for $\mu=m_*$.
Exact SPT curves are taken from Ref.~\cite{spt}.}
\label{p0+10}
\end{figure}

In Fig.~\ref{tadgap}, we show the 
truncated solutions $m_*(T)$ to the one-loop tadpole
gap equation as a function of
$g(2\pi T)$ for (a) $\mu=2\pi T$ and (b) $\mu=m_*$. 
The solutions have been
normalized to the leading order screening mass $m_{\rm LO}=g(2\pi T)T/\sqrt{24}$.
The truncated solutions were determined by treating $m$ as a quantity that is
${\cal O}(g)$ and truncating at a fixed order in $g$.  A $g^2$ truncation of
Eq.~(\ref{oneloopgapeq}), for example, yields $\hat{m}_*^2 = \alpha/6$,
which corresponds to the leading order screening mass.  The non-trivial
truncations, $g^3$ and $g^4$, are shown as grey dashed lines along with 
``exact'' curves from Ref.~\cite{spt} which are accurate to all orders in $m/T$.  
As can been seen from the figure, the gap equation converges very quickly
to the exact solutions for $\mu=2\pi T$ while for $\mu=m_*$ they do not
seem to be converging.  The primary difference between the two scales is
that in the case $\mu=m_*$ there are additional $\log\hat{m}$.  It is possible
that these logs need to be further resummed.  Note that as the renormalized
coupling constant becomes larger than $g(2\pi T)\sim4$ the uncertainty due to
the variation of the renormalization scale $\mu$ becomes rather large due
to the Landau singlarity present in the running of $g$.  For this reason,
in all results presented, we restrict ourselves to $g(2\pi T)\leq 4$.

\section{SPT-improved variables}
In this section, we use the solutions to the tadpole gap equation obtained in 
Sec.~\ref{tadders} to
obtain successive approximations to the pressure, screening mass, and
entropy
in screened perturbation 
theory.

\subsection{Pressure}

The two-loop SPT-improved
approximation to the pressure is obtained by inserting the
solution to the one-loop gap equation~(\ref{pet}) into the two-loop 
pressure~(\ref{2p}).
We can simplify the expression by using
Eq.~(\ref{pet}) to eliminate the explicit
appearance of logarithms of $\mu$.
This eliminates all the terms of order $\alpha$ and the expression
reduces to
\bqa
{\cal P}_{0+1} &=& 
{\cal P}_{\rm ideal} \Bigg[
1-{15\over2}\hat{m}^2
+15\hat{m}^3 \Bigg]
\;.
\label{P-01}
\eqa

\begin{figure}[ht]
\epsfxsize=15.5cm
\centerline{\epsffile{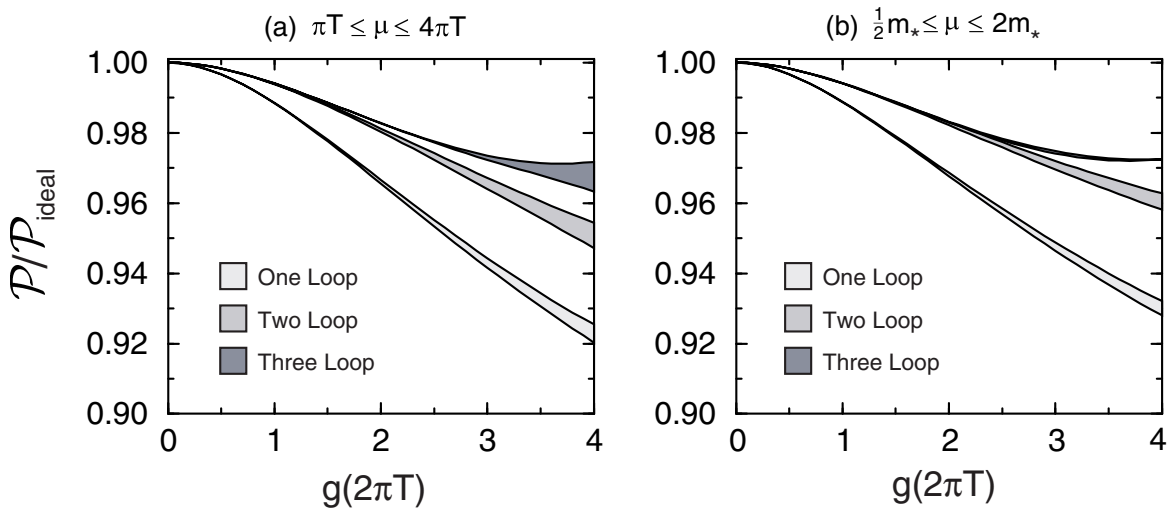}}
\vspace{3mm}
\caption[a]{The one-, two and three-loop SPT improved approximations to the
pressure as a function of $g(2\pi T)$ for (a) $\pi T < \mu < 4 \pi T$ and (b) ${1 \over 2} m_* < \mu < 2 m_*$.}
\label{pfig}
\end{figure}

In  Figs.~\ref{p0+1} and \ref{p0+10}, we show truncations of the 
one-, two-, and three-loop approximations
to the pressure for $\mu=2\pi T$ and $\mu=m_*$, respectively.  
The various truncations are shown as grey dashed lines along with
``exact'' curves from Ref.~\cite{spt} which are accurate to all orders in $m/T$.
As can be seen from the figure, the $m/T$ truncations converge very quickly
for the one- and two-loop approximations with the final two truncations
being virtually indistinguishable from the exact SPT solutions.
At three-loops, however, one needs to include
all terms up to $g^5$ before a reasonable approximation is obtained.  We 
therefore conclude that it is necessary to include higher order terms 
in order to fully converge to the exact SPT result at three-loops.  Also
it appears that the $m/T$ truncations converge better for $\mu=2\pi T$, than
for $\mu=m_*$.  Despite these caveats, at
all loop orders presented here, the highest order $m/T$ truncation provide an 
excellent approximation to the exact SPT results.

In Fig.~\ref{pfig} we show the
one-, two-, and three-loop approximations obtained using our $g^5$ truncation
to the pressure.  
The bands shown correspond to the results obtained by varying the renormalization scale $\mu$
over (a) $\pi T < \mu < 4\pi T$ and (b) ${1\over2}m_*<\mu<2m_*$.
This figure demonstrates that the $g^5$ truncations of the pressure 
yield a convergent series
of approximations which have very small variations with respect
to the renormalization scale.

\subsection{Screening Mass}

The one-loop SPT-improved approximation to the screening mass $m_s$ is given
by the solution to the tadpole gap equation~(\ref{tad}). 
A two-loop SPT-improved approximation can be obtained by inserting the 
solution to the gap equation for the mass parameter into Eq.~(\ref{scr2}).

\begin{figure}[ht]
\epsfxsize=15.5cm
\centerline{\epsffile{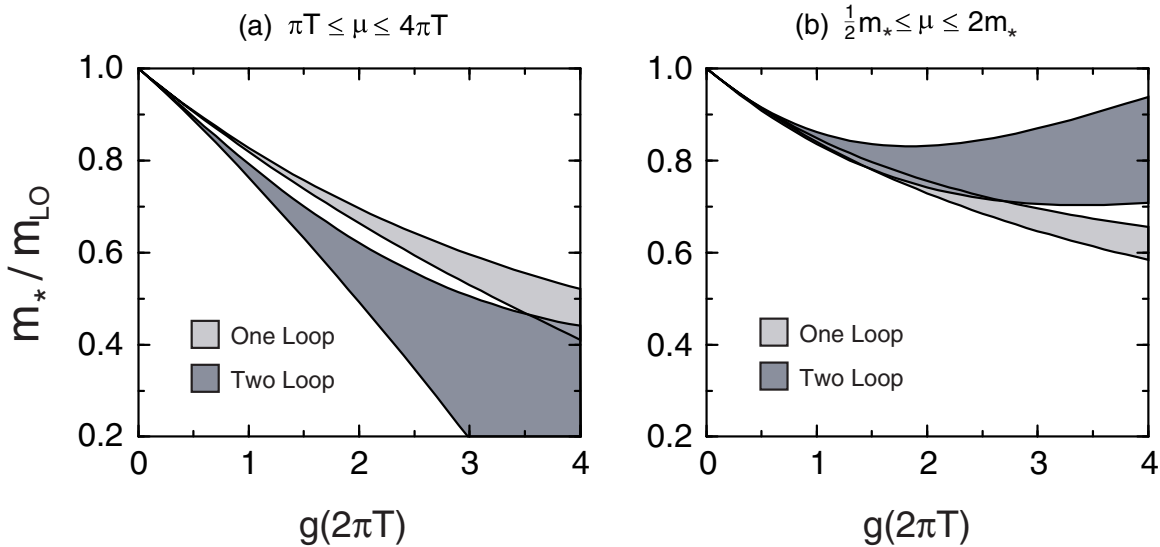}}
\vspace{3mm}
\caption[a]{The one- and two-loop SPT improved approximations to the
screening mass as a function of $g(2\pi T)$ 
(a) $\pi T < \mu < 4 \pi T$ and (b) ${1 \over 2} m_* < \mu < 2 m_*$.}
\label{sc0+10}
\end{figure}

In Fig.~\ref{sc0+10}, we show the $g^4$ truncations of the one- and two-loop approximations
to the screening mass.   The bands shown correspond to the results obtained 
by varying the renormalization scale $\mu$ over (a) $\pi T < \mu < 4\pi T$ 
and (b) ${1\over2}m_*<\mu<2m_*$.  One can see from this figure that the
convergence of the $m/T$ expansion for the screening mass is not as impressive
as for the pressure meaning that higher order truncations are necessary to
reliably describe the screening mass.
Also, we again see that the $m/T$ truncations converge better 
for $\mu=2\pi T$, than for $\mu=m_*$.

\subsection{Entropy}

The one, two-, and three-loop SPT-improved entropies are
obtained by replacing $m$ in the expressions~(\ref{s0})-(\ref{s2})
for ${\cal S}_0$, ${\cal S}_{0+1}$, and ${\cal S}_{0+1+2}$ with the solution 
to the one-loop gap equation.  As was the case with the two-loop pressure
we can use the gap equation to eliminate the logarithm $L$ yielding the
following expression for the two-loop entropy
\bqa
{\cal S}_{0+1}=
{\cal S}_{\rm ideal} \Bigg[
1 -{15\over2}\hat{m}^2T+15\hat{m}^3
-{45\over4}\hat{m}^4\Bigg]
\;.
\eqa
This is identical to the one-loop expression Eq.~(\ref{s0}), which is the
entropy of an ideal gas of particles with mass $m$.

\begin{figure}[ht]
\epsfxsize=15.5cm
\centerline{\epsffile{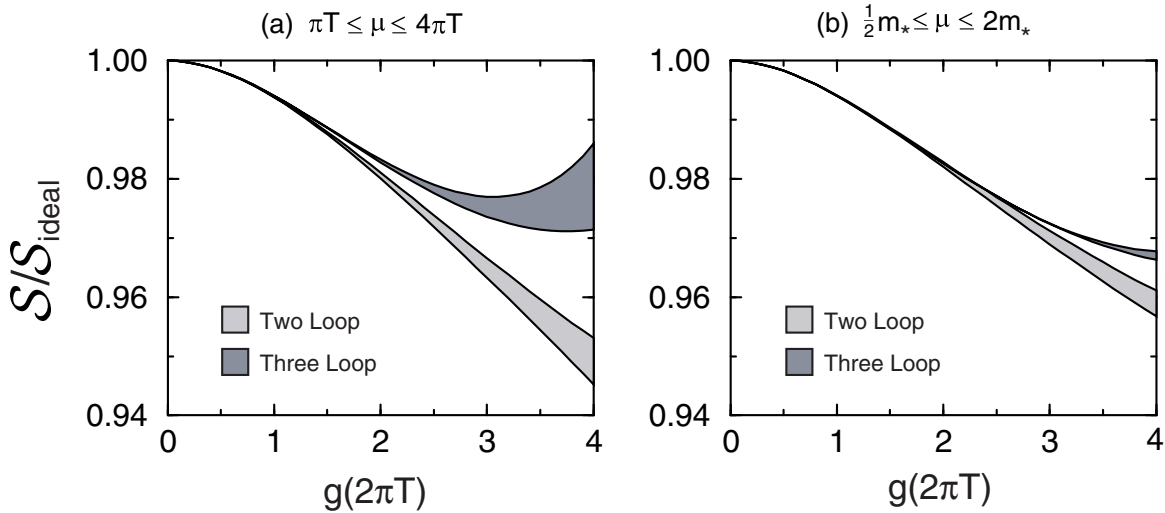}}
\vspace{3mm}
\caption[a]{
The SPT-improved two- and three-loop approximations to the
entropy as a function of $g(2\pi T)$
(a) $\pi T < \mu < 4 \pi T$ and (b) ${1 \over 2} m_* < \mu < 2 m_*$.} 
\label{s0+1}
\end{figure}

In Fig.~\ref{s0+1}, we show the ${\cal O}(g^5)$ truncations of the 
one-, two-, and three-loop approximations to the
entropy as a function of $g(2\pi T)$. 
The bands shown correspond to the results obtained 
by varying the renormalization scale $\mu$ over (a) $\pi T < \mu < 4\pi T$ 
and (b) ${1\over2}m_*<\mu<2m_*$.  In both cases the ${\cal O}(g^5)$ truncation
provides an excellent approximation to the exact SPT result.
Again we see that the $m/T$ truncations converge better 
for $\mu=2\pi T$, than for $\mu=m_*$.

\section{Conclusions}

In this paper, we have continued 
the systematic study of screened 
perturbation theory from Ref.~\cite{spt}. 
We applied it to the pressure and the entropy calculated
to three loops and the screening mass calculated to two loops.
By performing an expansion of the sum-integrals in powers of $m/T$, we
were able to obtain purely analytical results, without having to evaluate
integrals numerically.

Our calculations show that a truncation of the $m/T$ expansion
at ${\cal O}(g^5)$ is sufficient
to obtain accurate approximations to the exact one- and  two-loop results.
For the one- and two-loop approximations to the pressure and entropy, the
numerical results obtained in \cite{spt} and the truncated $m/T$ expansions  
are virtually indistinguishable as can be seen in Figs.~\ref{p0+1} and \ref{p0+10}.
At three-loop the ${\cal O}(g^5)$ truncation provides a reasonable description
of the pressure, but it seems that higher-order truncations are necessary to
provide accurate descriptions.  The fact that the $m/T$ expansions converge
quickly is important since
performing the ``exact'' SPT calculations is much more difficult than the
$m/T$ expansions.  An additional benefit of the $m/T$ expansion method is that
the final results can be determined completely analytically.

In Ref.~\cite{EJM1}, a generalization of SPT to gauge theories 
based on hard thermal loop (HTL) perturbation theory was proposed. The 
thermodynamic functions such as the pressure and entropy were calculated 
to one-loop order.  The two-loop calculation of the pressure
in QCD based on HTL perturbation theory requires not only
HTL propagators, but HTL vertices as well.  The exact calculation appears to be
very difficult.  Expansions like the one
presented here provide the simplification needed to complete the two-loop
HTL calculation~\cite{qcd2}.
The rapid convergence of the $m/T$ expansion in screened perturbation
theory is very encouraging in this regard.

\section*{acknowledgments}
The authors would like to thank E. Braaten and E. Petitgirard
for useful discussions
and suggestions.
This work was supported in part 
by the Stichting Fundamenteel Onderzoek der Materie
(FOM), which is supported by the Nederlandse Organisatie voor Wetenschapplijk
Onderzoek (NWO), and by 
the U.~S. Department of
Energy Division of High Energy Physics
(grant DE-FG03-97-ER41014).

\appendix
\renewcommand{\theequation}{\thesection.\arabic{equation}}
\section{Sum-integrals}
\label{appa}
In the imaginary-time formalism for thermal field theory, 
the 4-momentum $P=(P_0,{\bf p})$ is Euclidean with $P^2=P_0^2+{\bf p}^2$. 
The Euclidean energy $p_0$ has discrete values:
$P_0=2n\pi T$ for bosons and $P_0=(2n+1)\pi T$ for fermions,
where $n$ is an integer. 
Loop diagrams involve sums over $P_0$ and integrals over ${\bf p}$. 
With dimensional regularization, the integral is generalized
to $d = 3-2 \epsilon$ spatial dimensions.
We define the dimensionally regularized sum-integral by
\bqa
  \hbox{$\sum$}\!\!\!\!\!\!\int_{P}& \;\equiv\; &
  \left(\frac{e^\gamma\mu^2}{4\pi}\right)^\epsilon\;
  T\sum_{P_0=2n\pi T}\:\int {d^{3-2\epsilon}p \over (2 \pi)^{3-2\epsilon}}\;,
\label{sumint-def}
\eqa
where $3-2\epsilon$ is the dimension of space and $\mu$ is an arbitrary
momentum scale. 
The factor $(e^\gamma/4\pi)^\epsilon$
is introduced so that, after minimal subtraction 
of the poles in $\epsilon$
due to ultraviolet divergences, $\mu$ coincides 
with the renormalization
scale of the $\overline{\rm MS}$ renormalization scheme.

The one-loop sum-integrals that arise in the calculations 
have the following form
\bqa
{\cal I}^\prime_0&=&\sumint_{P} \log P^2 \;, \nonumber \\
{\cal I}_n&=&\sumint_{P}{1\over(P^2)^n}\;.
\eqa
Expanding in $\epsilon$ to the required order,
the specific one-loop sum-integrals needed are
\bqa
{\cal I}_0^{\prime}&=&-\left({\mu\over 4\pi T}\right)^{2\epsilon}
{\pi^2T^4\over45}\Bigg[1+{\cal O}(\epsilon)\Bigg]\;,
\\ 
{\cal I}_1
&=&
\left({\mu\over 4\pi T}\right)^{2\epsilon}{T^2\over12}
\left[1+\left(2+2{\zeta^{\prime}(-1)\over\zeta(-1)}\right)\epsilon
\right.\\
&&\left. \hspace{3.5cm}
+\left({\pi^2\over4}+4+4{\zeta^{\prime}(-1)\over\zeta(-1)}
+2{\zeta^{\prime\prime}(-1)\over\zeta(-1)}
\right)\epsilon^2 + {\cal O}(\epsilon^3)
\right]
\;,\\
{\cal I}_2
&=&{1\over(4\pi)^2}\left({\mu\over 4\pi T}\right)^{2\epsilon}
\left[{1\over\epsilon}+2\gamma+\left({\pi^2\over4}-4\gamma_1\right)\epsilon
+\left(
{\pi^2\over2}\gamma+4\gamma_2-{7\over3}\zeta(3)
\right)\epsilon^2 + {\cal O}(\epsilon^3)\right]
\;.
\eqa
The numbers $\gamma_1$ and $\gamma_2$ are the first and
second Stieltjes gamma constants
defined by the equation
\begin{equation}
\zeta(1+z) = {1 \over z} + \gamma - \gamma_1 z 
+{1\over2}\gamma_2 z^2
+ O(z^3)\;.
\end{equation}
The specific two-loop sum-integrals needed is
\bqa
\label{masslesssun}
\sumint_{PQ}{1\over P^2Q^2(P+Q)^2}
&=&0\;.\\ \nonumber
\eqa
It was first calculated by Arnold and Zhai in Ref.~\cite{arnold-zhai}.
The specific three-loop sum-integral needed is
\begin{eqnarray}
&& \sumint_{PQR}\frac{1}{P^2 Q^2 R^2 (P+Q+R)^2}
\nonumber
\\&& 
\hspace{3cm}  
\;=\; {T^4 \over 24(4\pi)^2} \left({\mu\over4\pi T}\right)^{6 \epsilon}
\left[ {1 \over \epsilon} + {91 \over 15} 
	+ 8 {\zeta'(-1) \over \zeta(-1)} - 2 {\zeta'(-3) \over \zeta(-3)}
	+ O(\epsilon) \right] \,.
\label{masslessbasket}
\end{eqnarray}
It was first calculated by Arnold and Zhai in Ref.~\cite{arnold-zhai}.

\section{Integrals}
\label{appb}
We also need some three-dimensional integrals. We choose dimensional
regularization to regulate infrared and ultraviolet divergences.
The integrals are generalized to 
$d=3-2\epsilon$ dimensions of space and $\mu$ is an arbitrary
momentum scale. 
\bqa
\int_{\bf p}&=&
  \left(\frac{e^\gamma\mu^2}{4\pi}\right)^\epsilon\;
\int {d^{3-2\epsilon}p \over (2 \pi)^{3-2\epsilon}}\;.
\eqa
The factor $(e^\gamma/4\pi)^\epsilon$
is introduced so that, after minimal subtraction 
of the poles in $\epsilon$
due to ultraviolet divergences, $\mu$ coincides 
with the renormalization
scale of the $\overline{\rm MS}$ renormalization scheme.

The integrals that arise in the calculations have the following form
\bqa
I^\prime_0&=&\int_{{\bf p}}\log(p^2+m^2) \;, \nonumber \\
I_n&=&\int_{{\bf p}}{1\over(p^2+m^2)^n}\;.
\eqa
Expanding in $\epsilon$ to the required order,
the specific one-loop integrals needed are
\bqa\
I_0^{\prime}
&=&
-{m^3\over6\pi}\left({\mu\over2m}\right)^{2\epsilon}\left[
1+{8\over3}\epsilon
+\left({\pi^2\over4}+{52\over9}\right)
\epsilon^2+ {\cal O}(\epsilon^3)
\right]
\;, \\
I_1
&=&
-{m\over4\pi}\left({\mu\over2m}\right)^{2\epsilon}\left[
1+2\epsilon
+\left({\pi^2\over4}+4\right)\epsilon^2 + {\cal O}(\epsilon^3)
\right]\;,\\
I_2&=&
{1\over8\pi m}\left({\mu\over2m}\right)^{2\epsilon}
\left[
1+{\pi^2\over4}
\epsilon^2 + {\cal O}(\epsilon^3)
\right]\;.
\eqa
The only two-loop integral needed is \cite{Braaten-Nieto:scalar}
\bqa\nonumber
I_{\rm sun} = \int_{\bf qr}{1\over q^2+m^2}{1\over r^2+m^2}{1\over({\bf p}+{\bf q}+{\bf r})^2+m^2}
\Bigg|_{{\bf p}^2=-m^2}
&=& \\
&&
\hspace{-2cm}
{1\over(8\pi)^2}
\left({\mu\over2m}\right)^{4\epsilon}
\left[
{1\over\epsilon}+6-8\log2
+{\cal O}(\epsilon)
\right]\;.
\label{3sun}
\eqa
The only three-loop integral needed is \cite{Braaten-Nieto:scalar}
\bqa\nonumber
I_{\rm ball} = \int_{\bf pqr}{1\over P^2+m^2}
{1\over q^2+m^2}{1\over r^2+m^2}{1\over({\bf p}+{\bf q}+{\bf r})^2+m^2}
&=& \\
&&
\label{Iball}
\hspace{-4cm}
-{m\over(4\pi)^3}\left({\mu\over2m}\right)^{6\epsilon}
\Bigg[
{1\over\epsilon}+8-4\log2+{\cal O}(\epsilon)
\Bigg]\;.
\eqa

\renewcommand{\theequation}{\thesection.\arabic{equation}}
\section{Setting Sun Diagram}
\label{appc}
The only nontrivial sum-integral required to calculate the self-energy
to two loops is the sunset diagram, which depends on the external
four-momentum $P=(p_0,{\bf p})$:
\bqa
\label{isundef}
{\cal I}_{\rm sun}(P)
=\sumint_{QR}{1\over Q^2+m^2}{1\over R^2+m^2}{1\over (P+Q+R)^2+m^2}\;.
\eqa
The sum-integral~(\ref{isundef}) must be evaluated at 
$p_0=0$ and ${\bf p}^2=-m^2$.
The setting-sun sum-integral 
involves a double sum-integral, so there are three 
momentum regions. The region where both $Q$ and $R$ are hard
is denoted by
$(hh)$, the region where one momentum is hard and the other
soft is denoted by $(hs)$, and the regions where both momenta are soft
is denoted by $(ss)$.
The contribution from each of these regions are
\bqa
{\cal I}_{\rm sun}^{(hh)}&=&
\sumint_{QR}{1\over Q^2R^2(Q+R)^2}
+{\cal O}(m^2)
\;,\\
\label{contri}
{\cal I}_{\rm sun}^{(hs)}&=&
{\cal O}(m)
\;,\\
\label{contri2}
{\cal I}_{\rm sun}^{(ss)}&=&
T^2 I_{\rm sun}
+{\cal O}(m^2)
\;.
\eqa

\renewcommand{\theequation}{\thesection.\arabic{equation}}

\section{Basketball Sum-integral}
\label{appd}

${\cal I}_{\rm ball}$ is the basketball sum-integral:
\begin{eqnarray}
{\cal I}_{\rm ball} &=& 
\sumint_{PQR} {1 \over P^2 + m^2} {1 \over Q^2 + m^2} 
	{1 \over R^2 + m^2} {1 \over S^2 + m^2} \,,
\label{IBall}
\end{eqnarray}
where $S = -(P+Q+R)$.  

The basketball sum-integral (\ref{IBall}) involves a triple
sum-integral, so there are 4 momentum regions:  $(hhh)$, $(hhs)$,
$(hss)$, and $(sss)$.  The contribution from each of these regions 
to order $m^2$ is
\begin{eqnarray}
{\cal I}_{\rm ball}^{(hhh)} &=&  
\sumint_{PQR} {1 \over P^2 Q^2 R^2 (P+Q+R)^2} 
+ {\cal O} (m^2),
\\
{\cal I}_{\rm ball}^{(hhs)} &=& 
4T I_1 \sumint_{QR} 
{1 \over Q^2 R^2 (Q+R)^2} 
+ {\cal O} (m^2),
\label{Iball-hss}
\\
{\cal I}_{\rm ball}^{(hss)} &=& 
{\cal O} (m^2),
\\
{\cal I}_{\rm ball}^{(sss)} &=& 
T^3 I_{\rm ball}
+ {\cal O}(m^2)
\;.
\end{eqnarray}
The $(hhh)$ contribution is given by  Eq.~(\ref{masslessbasket}), while
the $(hhs)$ contribution vanishes due to Eq.~(\ref{masslesssun}).
The (sss) contribution is given by Eq.~(\ref{Iball}).


\begin{thebibliography}{99}
\bibitem{arnold-zhai}
P.~Arnold and C.~Zhai, Phys. Rev. {\bf D50}, 7603 (1994);
	Phys. Rev. {\bf D51}, 1906 (1995);

\bibitem{Kastening-Zhai}
B.~Kastening and C.~Zhai, Phys.~Rev.~{\bf D52}, 7232 (1995).

\bibitem{Braaten-Nieto:QCD}
E.~Braaten and A.~Nieto, Phys.~Rev.~Lett.~{\bf 76}, 1417 (1996); 
       Phys.~Rev.~{\bf D53}, 3421 (1996).

\bibitem{Parwani-Singh}
R.R.~Parwani and H.~Singh, Phys. Rev. {\bf D51}, 4518 (1995).

\bibitem{Braaten-Nieto:scalar}
E.~Braaten and A.~Nieto, Phys. Rev. {\bf D51}, 6990 (1995).

\bibitem{Kleinert}
H.~Kleinert et al., Phys.~Lett.~{\bf B272}, 39 (1990); {\bf 319}, 545(E) (1993).

\bibitem{pade}
B.~Kastening, Phys.~Rev.~{\bf D56}, 8107 (1997);
T.~Hatsuda, Phys.~Rev.~{\bf D56}, 8111 (1997).

\bibitem{Baym}
G.~Baym, Phys.~Rev.~{\bf 127}, 1391 (1962).


\bibitem{Luttinger-Ward}
J.M.~Luttinger and J.C.~Ward, Phys.~Rev.~{\bf D118}, 1417 (1960).

\bibitem{comp1}
J.-P.~Blaizot, E.~Iancu, and A.~Rebhan, 
	Phys.~Rev.~Lett.~{\bf 83}, 2906 (1999); 
	Phys.~Lett.~{\bf B470}, 181 (1999).

\bibitem{BIR}
J.-P.~Blaizot, E.~Iancu, and A.~Rebhan, hep-ph/0005003 (2000).

\bibitem{K-P-P}
F.~Karsch, A.~Patk\'os, and P.~Petreczky, Phys. Lett. {\bf B401}, 69 (1997).

\bibitem{spt}
J.O.~Andersen, E.~Braaten and M.~Strickland, 
	Phys. Rev. {\bf D62}, 045004 (2000);
	Phys. Rev. {\bf D63}, 105008 (2001).

\bibitem{Chiku-Hatsuda}
S.~Chiku and T.~Hatsuda, Phys. Rev. {\bf D58}, 076001 (1998);
	hep-ph/9809215.

\bibitem{kastening} 
B.~Kastening, Phys.~Rev.~{\bf D54}, 3965 (1996).


\bibitem{EJM1} 
J.O.~Andersen, E.~Braaten and M.~Strickland, 
	Phys.~Rev. Lett.~{\bf 83}, 2139 (1999); 
	Phys.~Rev.~{\bf D61}, 14017 (2000); 
	Phys.~Rev.~{\bf D61}, 74016 (2000).
 
\bibitem{qcd2}
J.O.~Andersen, E.~Braaten, E.~Petitgirard and M.~Strickland, in preparation.
\end{thebibliography}
\end{document}